\documentclass[showpacs,preprint,preprintnumbers,showpacs,showkeys,superscriptaddress,amsmath,amssymb,nofootinbib]{revtex4}

\usepackage{amsmath}
\usepackage[dvips]{graphicx}
\usepackage[english]{babel}
\usepackage{wasysym}
\input psfig.sty
\begin{document}

\title{Probing the cosmological viability of non-gaussian statistics}

\author{Rafael C. Nunes}\email{rafadcnunes@gmail.com}
\affiliation{Departamento de  F\'isica, Universidade Federal de Juiz de Fora, 36036-330, Juiz de Fora - MG, Brasil}

\author{Ed\'esio M. Barboza Jr.}\email{edesiobarboza@uern.br}
\affiliation{Departamento de F\'isica, Universidade do Estado do Rio Grande do Norte, 59610-210, Mossor\'o - RN, Brasil}

\author{Everton M. C. Abreu}\email{evertonabreu@ufrrj.br}
\affiliation{Grupo de F\'isica Te\'orica e Matem\'atica F\'isica, Departamento de F\'isica,
Universidade Federal Rural do Rio de Janeiro, 23890-971, Serop\'edica-RJ, Brasil}
\affiliation{Departamento de  F\'isica, Universidade Federal de Juiz de Fora, 36036-330, Juiz de Fora - MG, Brasil}

\author{Jorge Ananias Neto}\email{jorge@fisica.ufjf.br}
\affiliation{Departamento de  F\'isica, Universidade Federal de Juiz de Fora, 36036-330, Juiz de Fora - MG, Brasil}

\begin{abstract}
\noindent 

Based on the relationship between thermodynamics and gravity we propose, with the aid of Verlinde's formalism, an 
alternative interpretation of the dynamical evolution of the Friedmann-Robertson-Walker Universe. This description takes into account 
the entropy and temperature intrinsic to the horizon of the universe due to the information holographically stored there 
through non-gaussian statistical theories proposed by Tsallis and Kaniadakis. The effect of these non-gaussian statistics in the cosmological context is change the strength of the gravitational constant. 
In this paper, we consider 
the $w$CDM model modified by the non-gaussian statistics and investigate the compatibility of these non-gaussian modification with the cosmological observations. In order to analyze in which extend the cosmological data constrain these non-extensive statistics, we use type Ia supernovae, baryon acoustic oscillations, Hubble expansion rate function and the linear growth of matter density perturbations data. 

\end{abstract}

\maketitle

\section{Introduction}
\label{intro}


There are theoretical evidences that the understanding of gravity has been greatly benefited 
from a possible connection to thermodynamics. Pioneering works of Bekenstein \cite{Bekenstein} and Hawking \cite{Hawking} 
have described this issue. For example, quantities as area and mass of black holes are associated with 
entropy and temperature respectively. Working on this subject, Jacobson \cite{Ted} interpreted Einstein field 
equations as a thermodynamic identity. Padmanabhan \cite{Padmanabhan} gave an interpretation of gravity
as an equipartition theorem. Recently, Verlinde \cite{Verlinde} brought an heuristic derivation
of gravity, both Newtonian and relativistic, at least for static spacetime. 
The equipartition law of energy has also played an important role. The analysis of the dynamics of
an inflationary universe ruled out by the entropic gravity concept was investigated in \cite{Easson}. On the other hand, one
can ask: what is the point of view of gravitational models coupled with thermostatistical theories and vice-versa?

The concept introduced by Verlinde is analogous to Jacobson's \cite{Ted} one, who proposed a thermodynamic 
derivation of Einstein's equations. The result has shown that the gravitation law derived by Newton 
can be interpreted as an entropic force originated by perturbations in the information ``manifold'' 
caused by the motion of a massive body when it moves away from the holographic screen.
An holographic screen can be understood as a storage device for information which is constituted by bits. 
Bits are the smallest units of information. Verlinde used this idea
together with the Unruh result \cite{Unruh} and he obtained Newton's second law. The idea of a entropic gravity/cosmology has been extensively investigated in different contexts, see \cite{varios,Tsallis_dark_energy} for recent results

Moreover, assuming the holographic principle together with the equipartition law of energy,
the Newton law of gravitation could be derived. The connection between nonextensive statistical theory and
the entropic gravity models \cite{Ananias,Everton1} make us to realize an arguably bridge between nonextensivity and gravity
theories. In this paper we propose an alternative interpretation of the dynamical evolution of the 
Friedmann-Robertson-Walker Universe (FRW) through non-gaussian statistical theories. We use the most recent observational data of Supernovae of Type Ia (SN Ia), Baryon Acoustic Oscillation (BAO), Hubble parameter and the growth function to investigate the cosmological consequences of such modifications through dark energy (DE) models. 

This paper is organized as follows. In section II we will make a brief review of the formulations concerning the non-gaussian of 
Tsallis and Kaniadakis statistics. In section III we will present the formalism of Verlinde and its consequences for the 
gravitational framework. Section VI we introduce the modified dynamic FRW universe. In section V we use  SN Ia, BAO, $H(z)$ and $f(z)$ data to constrain the non-gaussian statistics modifications
on the $w$CDM modified model. 
Lastly, Sec. VI briefly delivers our main conclusions and offers some final 
remarks. As usual, a zero subscript means the present value of the corresponding quantity.

\section{Non-gaussian statistics}

The objective of this section is to provide the reader with the main tools that will be used in the next sections. 
Although both formalisms are well known in the literature, these brief reviews can emphasize precisely that there is a 
connection between both ideas and that it was established recently \cite{Ananias}. 
The study of entropy has been an interesting task through recent years thanks to the
fact that it can be understood as a measure of information loss concerning the microscopic degrees 
of freedom of a physical system, when describing it in terms of macroscopic variables. 
Appearing in different scenarios, we can conclude that entropy can be considered as
a consequence of the gravitational framework \cite{Bekenstein,Hawking}. 
These issues motivated some of us to consider other alternatives to the standard Boltzmann-Gibbs (BG) theory in order to work
with Verlinde's ideas together with other subjects \cite{Everton1}.

\subsection{Tsallis statistics}

An important formulation of a nonextensive (NE) BG thermostatistics has been proposed by Tsallis \cite{Tsallis} 
in which the entropy is given by the formulation

\begin{equation}
\label{entropia_tsallis}
S_q=k_B \frac{1-\sum_{i=1}^W p_i^{q}}{q-1} \qquad  \Big(\sum_{i=1}^W p_i \,=\,1 \Big)  \,\,,
\end{equation}
where $p_i$ is the probability of the system to be in a microstate, $W$ is the total number of configurations and q, known in the current 
literature as Tsallis parameter or the nonextensive parameter, is a real parameter quantifying the degree of nonextensivity. The definition of
entropy (\ref{entropia_tsallis}) has, as motivation, to analyze multifractals systems and it also possesses the usual properties of positivity, 
equiprobability, concavity and irreversibility. It is important to note that Tsallis' formalism contains the BG statistics as 
a particular case in the limit $q \rightarrow 1$ where the usual additivity of entropy is recovered. Plastino and Lima \cite{Plastino}  
used a generalized velocity distribution for free particles \cite{Raimundo}

\begin{equation}
\label{q_distribuicao_velocidades}
f_q(v)=B_q \Big[ 1 - (1-q)\frac{mv^2}{2k_B T}    \Big]^{\frac{1}{1-q}},
\end{equation}
where $B_q$ is a dependent normalization constant, $m$ and $v$ is the mass and velocity of the particle, respectively. They have derived a 
nonextensive equipartition law of energy whose expression is given by

\begin{equation}
\label{q_equiparticao}
E_q=\frac{1}{5-3q} N k_B T\,\,,
\end{equation}
where the range of $q$ is $0 \leq q < 5/3$. For $q = 5/3$ (critical value) the expression of the equipartition law of energy, 
Eq. (\ref{q_equiparticao}), diverges. It is easy to observe that for $q = 1$, the classical equipartition theorem for each microscopic 
degrees of freedom is recovered.

\subsection{Kaniadakis statistics}

Kaniadakis statistics \cite{kaniadakis1}, also called $\kappa$-statistics, similarly to the TT formalism generalizes the standard BG 
statistics initially by the introduction of $\kappa$-exponential and $\kappa$-logarithm defined by

\begin{equation}
\label{k_expodencial}
\exp_{\kappa}(f)=(\sqrt{1+\kappa^2f^2} \,+\, \kappa f)^{\frac{1}{\kappa}}\,\,,  
\end{equation}

\begin{equation}
\label{k_log}
\ln_{\kappa}(f)=\frac{f^{\kappa}-f^{-\kappa}}{2 \kappa}\,\,,
\end{equation}
where the following operation being satisfied

\begin{equation}
\label{}
\ln_{\kappa}(\exp_{\kappa}(f))=\exp_{\kappa}(\ln_{\kappa}(f))=f\,\,.
\end{equation}

By Eqs. (\ref{k_expodencial}) and (\ref{k_log}) we can observe that the $\kappa$-parameter deforms the usual definitions of the exponential 
and logarithm functions.

The $\kappa$-entropy associated with this $\kappa$-framework is given by

\begin{equation}
\label{k_entropia}
S_{\kappa}(f)=- \int d^3p f \frac{f^{\kappa}-f^{-\kappa}}{2 \kappa},
\end{equation}
which recovers the BG entropy in the limit $\kappa \rightarrow 0$. It is important to mention here that the $\kappa$-entropy satisfied the 
properties of concavity, additivity and extensivity. Tsallis' entropy satisfies the property of concavity and extensivity but not additivity. 
This property is not fundamental, in principle. The $\kappa$-statistics has been successfully applied in many experimental fronts. 
As examples we can mention cosmic rays \cite{kaniadakis2}, quark-gluon plasma \cite{kaniadakis3}, kinetic models describing a gas 
of interacting atoms and photons \cite{kaniadakis4} and financial models \cite{kaniadakis5}.

The kinetic foundations of $\kappa$-statistics lead to a velocity distribution for free particles given by \cite{ze}

\begin{equation}
\label{k_distribuiao_velocidades_0}
f_{\kappa}(v)= \Bigglb(\sqrt{1+\kappa^2(\frac{mv^2}{2 k_B T})^2} -\kappa \frac{mv^2}{2 k_B T}\Bigglb)^{\frac{1}{\kappa}}.
\end{equation}

The expectation value of $v^2$ is given by

\begin{equation}
\label{k_distribuiao_velocidades}
<v^2>_{\kappa}= \frac{\int_0^\infty f_{\kappa} v^2 dv}{\int_0^\infty f_{\kappa} dv}. 
\end{equation}

Using the integral relation \cite{kaniadakis6}

\begin{equation}
\label{}
\int \,dx\, x^{r-1}\exp_{\kappa}(-x)= \frac{2 \mid\kappa\mid^{-r}}{1+r\mid\kappa\mid} \frac{ \Gamma(\frac{1}{2\mid\kappa\mid}-\frac{r}{2})}{\Gamma(\frac{1}{2 \mid\kappa\mid}+\frac{r}{2})} \,\Gamma(r)\,\,,
\end{equation}
we have that

\begin{equation}
\label{}
<v^2>_{\kappa}= \frac{2 k_B T}{m} \frac{1}{2 \kappa} \frac{1+ \frac{1}{2}\kappa}{1+ \frac{3}{2}\kappa}
               \frac{ \Gamma(\frac{1}{2 \kappa}-\frac{3}{4})}{\Gamma(\frac{1}{2\kappa}+\frac{3}{4})} 
               \frac{ \Gamma(\frac{1}{2 \kappa}+\frac{1}{4})}{\Gamma(\frac{1}{2\kappa}-\frac{1}{4})}.
\end{equation}

The $\kappa$-equipartition theorem is then obtained as

\begin{equation}
\label{E_kappa}
E_{\kappa}=\frac{1}{2} N \frac{1}{2 \kappa} \frac{1+ \frac{1}{2}\kappa}{1+ \frac{3}{2}\kappa}
               \frac{ \Gamma(\frac{1}{2 \kappa}-\frac{3}{4})}{\Gamma(\frac{1}{2\kappa}+\frac{3}{4})} 
               \frac{ \Gamma(\frac{1}{2 \kappa}+\frac{1}{4})}{\Gamma(\frac{1}{2\kappa}-\frac{1}{4})} k_B T.
\end{equation}

%

The range of $\kappa$ is $0 \leq \kappa < 2/3$. For $\kappa = 2/3$ (critical value) the expression of the equipartition
law of energy, Eq. (\ref{E_kappa}), diverges. For $\kappa = 0$, the classical equipartition theorem for each microscopic
degrees of freedom can be recovered.

\section{Verlinde's Formalism and the modified gravitational constant}

The formalism proposed by E. Verlinde \cite{Verlinde} derives the gravitational acceleration by using, basically, 
the holographic principle and the equipartition law of energy. This model considers a spherical surface as the holographic screen, with a particle of mass $M$
positioned in its center. A holographic screen can be imagined as a storage device for information. The number of bits (the term bit means 
the smallest unit of information in the holographic screen) is assumed to be proportional to the area $A$ of the holographic screen

\begin{equation}
\label{bits}
N=\frac{A}{l^2_p},
\end{equation}
where $A = 4 \pi r^2$ and $l_p = \sqrt{G \hbar / c^3}$. In Verlinde's formalism we assume that the total energy of the bits on the screen is given by the equipartition law of energy 

\begin{equation}
\label{equipartition}
E= \frac{1}{2}N k_B T.
\end{equation}

It is important to mention here that the usual equipartition theorem, Eq. (\ref{equipartition}), is derived from the usual BG thermostatistics. In a nonextensive thermostatistics scenario, the equipartition law of energy will be modified in a sense that a nonextensive parameter $q$ will be introduced in its expression. Considering that the energy of the particle inside the holographic screen is equally divided through all bits then we can write the equation

\begin{equation}
\label{ }
Mc^2=\frac{1}{2}N k_B T.
\end{equation}

Using Eq. (\ref{bits}) and the Unruh temperature formula \cite{Unruh}

\begin{equation}
\label{ }
k_B T=\frac{1}{2 \pi}\frac{\hbar a}{c},
\end{equation}
we are in a position to derive the (absolute) gravitational acceleration formula

\begin{equation}
\label{lei_de_newton }
a=\frac{l^2_p c^3}{\hbar} \frac{M}{r^2}=G \frac{M}{r^2}. 
\end{equation}

We can observe that from Eq. (\ref{lei_de_newton }) the Newton constant $G$ is just written in terms of the fundamental constants, 
$G=l^2_p c^3 / \hbar$.


As an application of NE equipartition theorem in Verlinde's formalism we can use the NE equipartition formula, i.e., Eq. (\ref{q_equiparticao}). 
Hence, we can obtain a modified acceleration formula given by \cite{Everton1}

\begin{equation}
\label{lei_de_newton_tsallis}
a=G_q \frac{M}{r^2},
\end{equation}
where $G_q$ is an effective gravitational constant which is written as

\begin{equation}
\label{G_tsallis}
G_q= \frac{5-3q}{2}G.
\end{equation}

From result (\ref{G_tsallis}) we can observe that the effective gravitational constant depends on the NE parameter $q$. 
For example, when $q = 1$ we have $G_q$ = $G$ (BG scenario) and for $q = 5/3$ we have the curious and hypothetical 
result which is $G_q = 0$. This result shows us that $q = 5/3$ is an upper bound limit when we are dealing with the 
holographic screen. Notice that this approach is different from the one demonstrated in \cite{BH}, where the authors 
considered in their model that the number of states is proportional to the volume and not to the area 
of the holographic screen.

On the other hand, if we use the Kaniadakis equipartition theorem, Eq. (\ref{E_kappa}), in Verlinde's formalism, 
the modified acceleration formula is given by

\begin{equation}
\label{G_kaniadakis_0}
G_{\kappa}= G_{\kappa} \frac{M}{r^2}\,\,,
\end{equation}
where $G_{\kappa}$ is an effective gravitational constant which is written as

\begin{equation}
\label{G_kaniadakis2}
G_{\kappa}=  2 \kappa \frac{1+ \frac{3}{2}\kappa}{1+ \frac{1}{2}\kappa}
                \frac{ \Gamma(\frac{1}{2 \kappa}+\frac{3}{4})}{\Gamma(\frac{1}{2\kappa}-\frac{3}{4})}
               \frac{ \Gamma(\frac{1}{2 \kappa}-\frac{1}{4})}{\Gamma(\frac{1}{2\kappa}+\frac{1}{4})} G.
\end{equation}

From result (\ref{G_kaniadakis2}) we can observe that the effective gravitational constant depends on the $\kappa$ parameter. 
For example, from the Gamma functions properties we have that 

\begin{equation}
\label{}
\lim_{\kappa \longrightarrow 0 } 
               \frac{ \Gamma(\frac{1}{2 \kappa}-\frac{3}{4})}{\Gamma(\frac{1}{2\kappa}+\frac{3}{4})} 
               \frac{ \Gamma(\frac{1}{2 \kappa}+\frac{1}{4})}{\Gamma(\frac{1}{2\kappa}-\frac{1}{4})} = 2\kappa.
\end{equation}

Then, using Eq. (\ref{G_kaniadakis2}) we obtain for $\kappa = 0$ that $G_{\kappa}=G$ (BG scenario). 

\section{Dark energy models through non-gaussian statistics}
\label{kaniadakis_dark_energy}

It was demonstrated in \cite{Everton1} that one modification in the dynamics of the FRW universe 
in NE Tsallis' statistics can be obtained simply by making the prescription $G \rightarrow G_{q} =(5 - 3q)G/2$ in the 
standard field equations. From this proposal in \cite{Tsallis_dark_energy} new cosmological constraints on the 
parameter $q$ were obtained. Analogouly to Tsallis' statistics, we can modify the Friedmann's equation in the Kaniadakis framework by making the prescription $G \rightarrow G_{\kappa}$, with $G_{\kappa}$ given by (\ref{G_kaniadakis2}). Thus, for a homogeneous and isotropic universe filled by perfect fluids, the equations of motion for non-gaussian statistics can be written as

\begin{equation}
\label{Friedmann_1}
 H^2 + \frac{k}{a^2}= \frac{8 \pi}{3} G_{q\,(\kappa)} \rho
\end{equation}
and

\begin{equation}
\label{Friedmann_2}
 \dot{H}+H^2 = -\frac{4 \pi}{3} G_{q\,(\kappa)} (\rho +3p),
\end{equation}
where $H = \dot{a}/a$ is the Hubble function, $G_{q\,(\kappa)}$ denotes the effective gravitational constant in the Tsallis (Kaniadakis) formalism and $\rho$ and $p$ are, respectively, the total density and pressure of the fluid. These equations can be combined to obtain the conservation equation,

\begin{equation}
\label{eq_conservacao}
\dot{\rho} + 3H(1+w)\rho=0.
\end{equation}

For a FRW universe pervaded by radiation, non-relativistic matter (baryonic plus dark matter) and some sort of dark energy to take into account the late time cosmic acceleration, the Friedmann equation (\ref{Friedmann_1}) becomes

\begin{equation}
\label{Hz}
 \frac{H^2(a)}{H^2_0} = \frac{G_{q\,(\kappa)}}{G}\Big(\Omega_{\gamma,0}a^{-4}+\Omega_{m,0}a^{-3}+\Omega_{k,0}a^{-2} + \Omega_{x0}f(a) \Big)
\end{equation}
where $\Omega_{i,0} = 8 \pi G \rho_{i0} /(3H_0^2)$ is the density parameter of the $i$-th component ($i=\gamma,\,m,\,{\rm and}\, x$ for radiation, matter (baryonic 
more dark) and dark energy, respectively), $\Omega_{k,0}=-k/H_0^2$ is the curvature density parameter and

\begin{equation}
\label{dark_energy_evolution}
f(a)=\frac{\rho_x}{\rho_{x0}}=a^{-3}\exp \Big( -3 \int_1^a \frac{w(a') da'}{a'}  \Big)
\end{equation}
is the density ratio for a dark energy fluid with a generic equation of state parameter $w(a) \equiv p_x /\rho_x$. In the above equations the subscript $0$ denotes the observable at present time. 

In the non-gaussian scenario, the normalization condition reads as

\begin{equation}
\label{normalization}
\Omega_{\gamma,0} +\Omega_{m,0} + \Omega_{k,0} + \Omega_{x0}= \frac{G}{G_{q\,(\kappa)}},
\end{equation}
which is an interesting result since it can show us, one more time, that the value $q=1$ ($\kappa=0$) 
recovers the standard normalization condition. Values of $q>5/3$ and $\kappa < 0$ which brings a negative normalization condition 
makes no sense. 
\\ \

In this paper we assume spatial flatness  and we restrict ourselves to the case $w = const$ so that Eq. (\ref{Hz}) becomes


\begin{eqnarray}
\label{Hz_LCDM}
 \frac{H^2(a)}{H^2_0} &=& \frac{G_{q\,(\kappa)}}{G} \Big[\Omega_{\gamma,0}a^{-4}+\Omega_{m,0}a^{-3}+\Big(\frac{G}{G_{q\,(\kappa)}}-\Omega_{\gamma,0}-\Omega_{m0} \Big)(1+z)^{3(1+w)}\Big].
\end{eqnarray}

From Eqs. (\ref{G_tsallis}), (\ref{G_kaniadakis2}) and (\ref{normalization}) it is possible to note that the parameters $q$ and $\kappa$ affects the energy balance of the universe. If $q$ is greater (smaller) than one, the effective gravitational constant is smaller (grater) than $G$ so that more (less) dark energy will be required to provide the observed late time universe acceleration. By its turn, the Kaniadakis framework, if $\kappa > 0$, the gravitational field is weaker than in the gravitational field in the standard BG scenario so that we will need
more dark energy to accommodate the cosmic acceleration. Since $\kappa\ge0$, the Kaniadakis statistics is more restrictive than Tsallis statistics. In the next section, we use some of available cosmological observations to obtain new constraints on the non-gaussian statistical parameters $q$ and $\kappa$.

\section{Observational Constraints}

In order to constrain the parameters (q,$\kappa$ and $w$) we perform a joint analysis involving the $580$ SNe Ia distance  measurements of the Union2.1 data set ~\cite{SNIa}, the $30$ measurements of the Hubble parameter $H(z)$ given in Table $4$ of Ref.~\cite{Moresco_H}, $17$ measurements of the growth function $f(z)$ listed in Table I and the six estimates of the BAO parameter given in Table 3 of Ref.~\cite{BAO}. 

\subsection{Type Ia Supernovae}

The SNe Ia sample of the Union 2.1 is given in terms of distance modulus $ \mu$. Theoretically, the distance modulus is given by

\begin{equation}
\mu_i^{th}(z_i) = 5 \log{H_0d_L(z_i)} + 5\log(3/h)+40,
\end{equation}
where $h=H_0/100\,{\rm Km}\cdot{\rm s}^{-1}\cdot{\rm Mpc}^{-1}$ and

\begin{equation}
d_L(z)= (1+z)  \int_0^z  \frac{ dz'}{H(z')}
\end{equation}
is the luminosity distance for a spatially flat universe. The usual $ \chi^2$ function is calculated as

\begin{equation}
\label{chi}
\chi^2_{SN} = \sum_i \frac{\left( \mu^{th}_i- \mu^{obs}_i\right)^2}{\sigma^2_{\mu_i}},
\end{equation}
where $\mu_i^{obs}$ is the observed value of the distance modulus at redshift $z_i$ and $\sigma_{\mu_i}$ its uncertainty.
In our analysis we treat the Hubble constant $H_0$ as a nuisance parameter and marginalize over $H_0$ so that the $\chi^2$ function to be minimized is

\begin{equation}
\label{chi2}
\chi^2_{SN} = A-\frac{B^2}{C},
\end{equation}

\noindent where the quantities $A$, $B$, $C$ are given by:
\begin{equation}
A= \left( \mu^{th}_i- \mu^{obs}_i\right)(C^{-1}_{SN})_{ij}\left( \mu^{th}_j- \mu^{obs}_j\right),
\end{equation}
 \begin{equation}
B=\sum_i (C^{-1}_{SN})_{ij}\left( \mu^{th}_j- \mu^{obs}_j\right)
\end{equation}
and
\begin{equation}
C= \sum_{ij}(C^{-1}_{SN})_{ij}
\end{equation}
and $(C^{-1}_{SN})_{ij}$ is the inverse convariance matrix. 

\subsection{Growth function}

In the linear regime the matter density perturbations $\delta=\delta\rho_m/\rho_m$ satisfies
\begin{equation}
\label{density_pert_eq}
\ddot{\delta}+2H\dot{\delta}-4\pi G_{eff}\rho_m\delta=0,
\end{equation}

\noindent where $G_{eff}$ is the effective gravitational constant for a given theory of gravity. For the models studied in this paper, $G_{eff}$ is given by (\ref{G_tsallis}) for Tsallis statistics and by (\ref{G_kaniadakis2}) for Kaniadakis statistics. By defining the growth factor $f\equiv d\ln\delta/\ln a$, this second order time differential equation is reduced to 

\begin{equation}
\label{growth_factor_a}
f'+f^2+\Big(2-\frac{3}{2}\Omega_m\Big)f-\frac{3}{2}\Omega_m^{eff}=0,
\end{equation}

\noindent where $f'=df/d\ln a$ and

\begin{eqnarray}
\Omega_m^{eff}=\frac{8\pi G_{eff}\rho_m}{3H^2}=\frac{\Omega_{m,0}a^{-3}}{\Big[\Omega_{\gamma,0}a^{-4}+\Omega_{m,0}a^{-3}+\Big(\frac{G}{G_{q\,(\kappa)}}-\Omega_{\gamma,0}-\Omega_{m0} \Big)a^{-3(1+w)}\Big]}.
\end{eqnarray}

\noindent In Table I we list the $20$ measurements of $f$. The usual $ \chi^2$ function is calculated as

\begin{equation}
\label{chi_square_f}
\chi^2_f=\sum_{i=1}^{20}\frac{(f_i^{obs}-f_i^{theo})^2}{\sigma_{f_i}^2},
\end{equation}
where $f_i^{obs}$ is the observed value of the growth function at redshift $z_i$, $\sigma_{f_i}$ its uncertainty and $f_i^{theo}$ the value of $f(z_i)$ provided theoretically. In order to obtain $f_i^{theo}$ we use the approximation $f(z)\approx\Omega_m^{eft}(z)^{\gamma}$ \cite{wang}, where $\gamma$ is the growth index and depends of the underlying cosmological model. For the $w$CDM model, $\gamma=3(w-1)/(6w-5)$. 

\begin{table}
\caption{Currently available data for growth rates used here.}
\label{f_data}
\begin{tabular}{cccl}
\hline
\hline
$z$ & $f$ & $\sigma_f$ & Ref.\\
\hline
0.02 & 0.482&0.09&\cite{davis}\\
0.067&0.56&0.11&\cite{beutler}\\
0.11 & 0.54&0.21&\cite{bielby}\\
0.15 & 0.49& 0.14 & \cite{guzzo}\\
0.15 & 0.51 & 0.11 & \cite{verde}; \cite{hawkins}\\
0.22 & 0.60 & 0.10 & \cite{blakeb}\\
0.32 & 0.654 & 0.18 & \cite{reyes}\\
0.34 &0.64&0.09&\cite{cabre}\\
0.35 & 0.70 & 0.18 & \cite{tegmark}\\
0.41 & 0.70 & 007 & \cite{blakeb}\\
0.42 & 0.73 & 0.09 &\cite{blakec}\\
0.55 & 0.75 & 0.18 & \cite{ross}\\
0.59 & 0.75 & 0.09 &\cite{blakec}\\
0.60 & 0.73 & 0.07 &\cite{blakeb}\\
0.77 & 0.91 & 0.36 &\cite{guzzo}\\
0.78 & 0.70 & 0.08 &\cite{blakeb}\\
1.4 & 0.90 & 0.24 &\cite{dangela}\\
2.125 & 0.78 & 0.24&\cite{viel}\\
2.72 & 0.78 & 0.24 &\cite{viel}\\
3.0& 0.99&0.24&\cite{bielby}\\
\hline
\hline
\end{tabular}
\end{table}

\subsection{Hubble parameter} 

In Ref. \cite{Jimenez}, Jimenez has developed a method to use the relative age of old and passive galaxies, $dz/dt$, to infer the Hubble parameter as a function of the redshift,
\begin{equation}
H(z)=\frac{\dot a}{a}=-\frac{1}{1+z}\frac{dz}{dt}.
\end{equation}
The {\it cosmic chronometers} method to measure $H(z)$ does not depend of any integrated distance measurement over redshifts and is independent of cosmological models. We use $30$ $H(z)$ data obtained from the cosmic chronometers method listed in Table $4$ of \cite{Moresco_H}. The usual $ \chi^2$ function is calculated as

\begin{equation}
\label{chi_square_H}
\chi^2_H=\sum_{i=1}^{30}\frac{(H_i^{obs}-H_i^{theo})^2}{\sigma_{H_i}^2},
\end{equation}
where $H_i^{obs}$ is the observed value of the Hubble parameter at redshift $z_i$, $\sigma_{H_i}$ its uncertainty and $H_i^{theo}$ the value of the Hubble parameter at $z_i$ provided theoretically.

\subsection{BAO}

Our BAO analysis is based on the BAO parameter: 
\begin{equation}
\label{BAO_parameter}
{\cal{A}} (z) = D_V(z){\sqrt{\Omega_{{\rm m},0}^{q(k)} H_0^2}},
\end{equation}
where
\begin{equation}
D_V(z) = \Big[\frac{1}{H(z)}\Big(\frac{1}{z}\int_0^z\frac{dz'}{H(z')}\Big)^2\Big]^{1/3}
\end{equation}
is the so-called dilation scale and $\Omega_{{\rm m},0}^{q(k)}=G_{q(k)}\Omega_{{\rm m},0}$ is the matter density modified to take into account the effects of the non-gaussian statistics. Here we use the six estimates of the BAO parameter given in Table 3 of Ref.~\cite{BAO}. The usual $ \chi^2$ function is calculated as

\begin{equation}
\label{chi_square_BAO}
\chi^2_{{\rm BAO}}=\sum_{i=1}^{6}\frac{(A_i^{obs}-A_i^{theo})^2}{\sigma_{A_i}^2},
\end{equation}
where $A_i^{obs}$ is the observed value of the BAO parameter at redshift $z_i$, $\sigma_{A_i}$ its uncertainty and $A_i^{theo}$ the theoretical value of the BAO parameter at redshift $z_i$.

As the likelihood function is defined by $L\propto\exp(-\chi^2/2)$ the said values follow from minimize the quantity $\chi^2_{{\rm total}}=\chi^{{\rm SN Ia}}+\chi^2_{H}+\chi^2_{f}+\chi^2_{{\rm BAO}}$. In the following we use $\Omega_{{\rm m},0}=0.27$. 

\subsection{Results}
 \begin{figure}[!t]
 \includegraphics[width=4in, height=6in,angle=270]{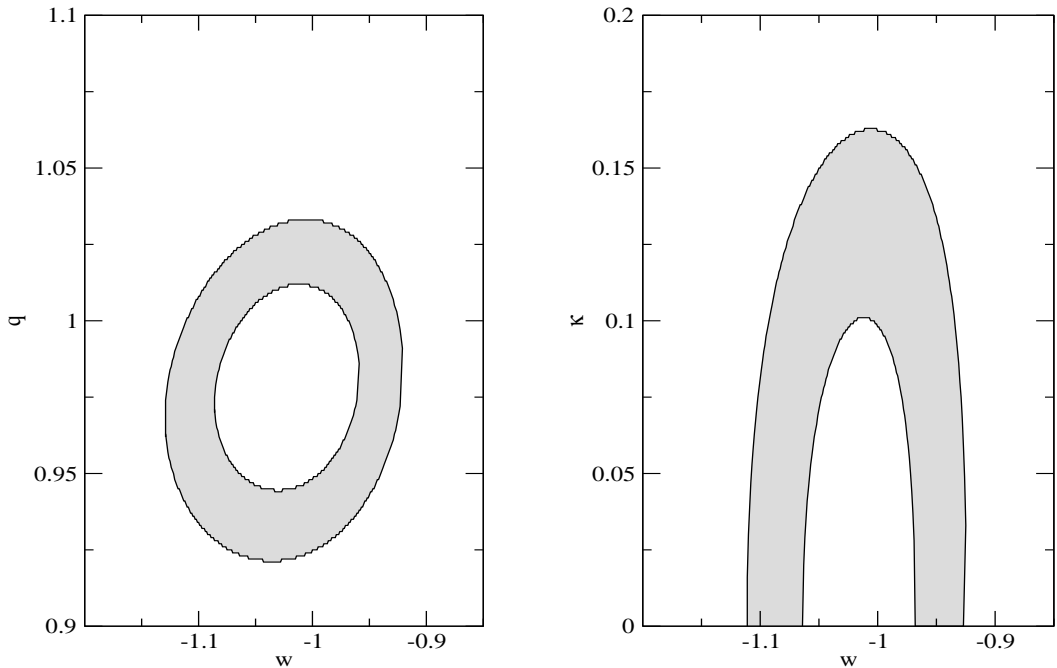}
 \caption{\label{ngps}  1$\sigma$ and 2$\sigma$ confidence regions for the modified $w$CDM model 
 from a joint analysis with SNIa + $f$ + $H(z)$ +BAO data. Left: Tsallis' statistics. Right: Kaniadakis Statistics.}
 \end{figure}


Figure \ref{ngps} shows the 68$\%$ and 95$\%$ confidence regions in the $q-w$ plan (left panel) and in the $\kappa-w$ plan (right panel) for
the modified $w$CDM model. At $2\sigma$, the best fit points are: $(q=0.98_{-0.046}^{+0.045},\,w= -1.02_{-0.086}^{+0.081})$  with  $\chi^2_{{\rm min}}=588.32$ and $(\kappa=0.00_{-0.000}^{+0.150},\,w=-1.01_{-0.083}^{+0.079})$ with $\chi^2_{{\rm min}}=589.25$. Quintessential ($w> - 1$) and phantom ($w<-1$) fluids are compatible with both, Tsallis and Kaniadakis statistics. As we can see, the cosmological observations used in this work are completely compatible with non-gaussian statistics. However, it is important to note that these data does not exclude the BG statistics, i. e., $q=1$ and $\kappa=0$. Tsallis' statistics provide a better fit to the data than Kaniadakis. For Tsallis' statistics $0\le G_{eff}/G\le 2.5$ while for Kaniadakis' statistics $0\le G_{eff}/G\le1$. Thus, we can to attribute this result to the additional freedom allowed by Tsallis' statistics since the data seems to choose a slightly greater gravitational constant. 

\section{Conclusion}

Currently, two extensions of standard statistical mechanics, known as Kaniadakis and Tsallis statistics, 
has been used to explain a very large class of phenomena observed experimentally in different areas, e.g, 
in low and high energy physics, astrophysics, econophysics, biology, ect.
In this paper, we have explored the possibility of one modification in the dynamics of the FRW universe obtained through an entropic 
force theory generalized for the Kaniadakis and Tsallis statistics.

From the combination SNIa+BAO+$H(z)+f(z)$ datasets 
we have obtained new cosmological constraints over $q$ and $\kappa$ parameters.


To sum up, in this paper, we have obtained new values for the free parameters that characterizes 
the non-gaussian statistical theory proposed by Kaniadakis and Tsallis.
Based in the data used in this paper, we note that non-gaussian statistics can not be rule out by cosmological observations although the BG statistics remains in fully agreement with the data.
From the results obtained here we can conclude that the parameters $q$ and $\kappa$ affects the energy balance of the universe. 
The gravitational field is more weak for $q>1$ and $\kappa > 0$ so that we need
more dark energy than we would have if we consider the standard BG scenario. For $0\leq<1$ the strength of the gravitational field greater than in the standard BG scenario and less dark energy is required to explain the cosmological observations. The results obtained in this paper favor a slightly strong gravitational field.
\acknowledgments

\noindent The authors thank CNPq (Conselho Nacional de Desenvolvimento Cient\' ifico e Tecnol\'ogico),
Brazilian scientific support federal agency, for partial financial support, Grants
numbers 302155/2015-5, 302156/2015-1 and 442369/2014-0 and E.M.C.A. thanks the hospitality of Theoretical Physics Department at Federal University of Rio de Janeiro (UFRJ), where part of this work was carried out.


\end{document}